\documentstyle[12pt,aaspp4]{article}

\def\kms{\ifmmode{\,\hbox{km}\,s^{-1}}\else {\rm\,km\,s$^{-1}$}\fi}

\def\ms{{\rm\,M_\odot}}
\def\lsun{{\rm\,L_\odot}}

\def\cm{{\rm\,cm}}

\def\kmps{{\rm\,km\,s$^{-1}$}~}
\def\kmpsnospace{{\rm\,km\,s$^{-1}$}}
\def\hmpc{\ifmmode{h^{-1}\,\hbox{Mpc}}\else{$h^{-1}$\thinspace Mpc}\fi}

\def\eg{{\it e.g.}~}
\def\etal{{\it et~al.}~}

\def\ie{{\it i.e.}~}

\def\spose#1{\hbox to 0pt{#1\hss}}
\def\lta{\mathrel{\spose{\lower 3pt\hbox{$\mathchar"218$}}
\raise 2.0pt\hbox{$\mathchar"13C$}}}
\def\gta{\mathrel{\spose{\lower 3pt\hbox{$\mathchar"218$}}
\raise 2.0pt\hbox{$\mathchar"13E$}}}

\def\H2{$H_2$~}
\def\H2S{$H_2^*$~}
\def\Mco{{$M_{H_2}$~}}

\def\D25{{$D_{25}$~}}

\def\HIH2LB{{$M_{GAS^*}/L_B$~}}

\def\MH2D2{{$M_{H_2}/D_{25}^2$~}}
\def\MH2D2S{{$M_{H_2^*}/D_{25}^2$~}}

\def\thCO{{$^{13}$CO~}}
\def\TA{{$T_A^*$}}
\def\Tk{{$T_k$~}}
\def\XdVunit{{\rm\,pc\,{[km\,$s^{-1}$]}$^{-1}$}~}
\def\cmsq{{$cm^{2}$}}
\def\um{{$\mu$m~}}
\def\1213CO{{[$^{12}$CO}/{$^{13}$CO]~}}
\def\nh2{{$n(H_2)$}}
\def\cm3{{$cm^{-3}$}}
\def\H2{{$H_2$}}

\begin{document}


\title
{A Multi-transition CO Study of The ``Antennae'' Galaxies NGC~4038/9 }

\author{Ming Zhu\altaffilmark{1}, E. R. Seaquist\altaffilmark{1}, and Nario Kuno\altaffilmark{2}}

\altaffiltext{1}{Dept. of Astronomy, U. of Toronto, 60 St. George St.
Toronto, ON M5S$-$3H8, Canada}

\altaffiltext{2}{Nobeyama Radio Observatory, Minamimakimura, Nagano 384-1305, Japan}

\begin{abstract}

For the Antennae interacting galaxy pair, we have obtained high quality, fully
sampled $^{12}$CO J=1--0 and 3--2 maps of the regions surrounding the nuclei
and the area of overlap between the two galaxies. The maps possess an
angular resolution of 15$''$ or 1.5 kpc, so far the highest resolution maps
available at both the J=1--0 and 3--2
transitions. In addition, $^{12}$CO J=2--1 data have been obtained for the 
positions of the two nuclei
as well as in part of the overlap region with 20$''$ angular
resolution. The $^{12}$CO J=1--0, 2--1, 3--2 emission all peak in
an off-nucleus region adjacent to where the two disks overlap. 
Use of the conventional $ X $ 
factor yields $ \sim 4 \times 10^9 \ms$ molecular gas
mass in the overlap region. It is difficult to understand how such a large amount 
of molecular gas can be accumulated in this region
given the relatively short lifetime of molecular clouds and the 
limited period of time for this region to form.

Line emission at $^{13}$CO J=2--1 and 3--2 is detected at selected
points in the two nuclei and the overlap region.  Both the
$^{12}$CO/$^{13}$CO J=2--1 and 3--2 integrated intensity ratios are
remarkably high in the overlap region. This is the first published
case in which such high $^{12}$CO/$^{13}$CO J=2--1 and 3--2 ratios are
found outside a galactic nucleus.  Detailed LVG modeling indicates
that the $^{12}$CO and \thCO emission originate in different spatial
components. The $ ^{12}$CO emission may originate within a non-virialized
low density gas component with a large velocity gradient. Assuming a
CO-to-H$_2$ abundance ratio of $10^{-4}$,
the X factor given by the LVG model is an order of magnitude lower than the
conventional value for molecular clouds in the Milky Way, but it scales 
inversely as the assumed value for this ratio. Accordingly, we 
suggest the possibility that the strong CO emission
in the overlap region of the Antennae galaxies is associated with increased
radiative efficiency, possibly caused by a large velocity dispersion within
the individual molecular clouds.

A comparison of the CO J=3--2 emission with the SCUBA 850 \um
continuum in the Antennae galaxies shows that the CO line emission 
contributes globally 46\% of the 850 \um continuum flux and that the ratio
of $^{12}$CO J=3--2 to SCUBA 850 \um flux varies by a factor of two
across the system.  After correcting for the $^{12}$CO J=3--2
contamination, the dust emission at 850 \um detected by SCUBA is
consistent with the thermal emission from a single warm dust component
with a mass of 1.7 $\times 10^7 \ms$.

\end{abstract}



\section{Introduction}

 The remarkable antennae galaxy pair
NGC~4038/9 is one of the nearest merging systems, and as such it has
been well studied at essentially all available wavelengths from radio to X-ray 
(\eg  Hummel \& van der Hulst 1986; Haas \etal 2000; Hibbard \etal 2001; Bushouse, Telesco \& Werner 1998;
Nikola \etal 1998; Mirabel \etal 1998; Vigroux \etal 1996;
Fabbiano, Schweizer \& Mackie 1997).  Its optical morphology, such as the tidal tail features,
has been successfully reproduced by N-body simulations (\eg Barnes 1988; Mihos, Bothun \& Richstone
1993;  Dubinski, Mihos \& Hernquist 1996). 
Therefore it is an ideal object for a detailed study of the
correlation between the interstellar medium (ISM) and the star
formation activity in mergers.
\nocite{Hummel86}
\nocite{Mirabel98}
\nocite{Bushouse98}
\nocite{Nikola98}
\nocite{Vigroux96}
\nocite{Mihos93}
\nocite{Dubinski96}
\nocite{Gao01}
\nocite{Young95}
\nocite{Bloemen86}
\nocite{Mauersberger99}
\nocite{Taniguchi98}
\nocite{Braine93}
\nocite{Aalto91}
\nocite{Aalto97}
\nocite{Goldreich74}
\nocite{Klaas97}
Of particular interest about  this archetypical early-stage merging
system is that the emission of CO, mid-IR (MIR), far-infrared (FIR), submillimeter (sub-mm) and radio
6 cm and 20 cm continuum all peak in an off-nucleus region which is
coincident with the overlap between the two colliding disks. The
recent release of HST images (Whitmore \etal 1999) have revealed a
population of hundreds of compact young clusters, whose ages are
estimated to be less than 10 Myr, distributed along the edge of a dust
lane in the overlap region. MIR~(Mirabel \etal 1998; Vigroux
\etal 1996), FIR~(Evans, Harper, \& Helou 1997; Bushouse
\etal 1998) and sub-mm (Haas \etal 2000) 
\nocite{Haas2000}
\nocite{whitmore99}
\nocite{Evans97}
images of this region
indicate even more intense starbursts that are heavily obscured by
dust and therefore inconspicuous at optical wavelengths. Huge amounts
of molecular gas have been found in the whole system as well as in the
overlap region (Gao \etal 2001). Assuming a conventional CO-to-H$_2$ 
conversion factor (the $X$ factor), the global star formation efficiency
$L_{\rm IR}/M_{H_2}$, however, is only $\sim 3.7 \lsun/\ms$, which
is similar to that of giant molecular clouds (GMCs) in our Galaxy. All
these suggest that the bulk of the molecular gas in this system is
probably still in an initial turning-on stage of the starburst. This
provides us with an excellent opportunity to study the impact of 
galaxy interactions and induced starbursts  on the molecular gas properties.

The main purpose of our study of the Antennae galaxies is to
determine the physical properties of the molecular gas in
the starbursting overlap region and the two galactic nuclei. 
This study will permit the density and temperature to be determined for the star forming regions
and a more reliable estimate of the mass of the molecular gas in these regions.

We describe a detailed excitation analysis using high resolution
multi-line CO data. This analysis is based on fully sampled maps at a common resolution of 15$''$
(1.5 kpc) so that line ratios can
be estimated reliably without the frequently employed ad hoc
assumptions about the source size and brightness distribution. The transitions used are CO J=1--0
and J=3--2 with both the $ ^{12}$CO and $ ^{13}$CO isotopes represented at J=2--1 and J=3--2
at selected  points, permitting us to examine a wide
range of molecular gas excitation conditions,  the $X$ factor and the molecular gas mass.
Thermal dust emission at sub-mm continuum wavelengths 
provides an estimate
of the dust mass. Our CO J=3--2 maps permit us to correct more precisely the 
850 \um map by Haas \etal 
(2000) for the contribution by this transition to arrive at a revised estimate of the dust mass.

\section{$ ^{\bf 12}$CO  J=1--0 Observations with the NRO 45m telescope}

 The $ ^{12}$CO J=1--0 observations at 115.27 GHz were conducted in two observing
 runs on 1998 March 23--26 and 1999 January 17--18 using the 45m telescope
of the Nobeyama Radio Observatory (NRO) in Japan.  The 2 $\times$ 2 multi-beam SIS receiver
S115Q was used, which provides relatively fast mapping of the
extended CO emission in NGC~4038/9. This receiver has four beams
separated by 34$''$ and with FWHM beamwidth of 15$''$ at the
frequency of CO J=1--0. The system temperature during these observations
was in the range 350--700 K. The wide band acousto-optical spectrometer
(AOS) was used as the back end which provides a total usable
bandwidth of 250 MHz and a frequency resolution of 250 KHz. At
115GHz, these correspond to a total velocity coverage of 650 \kmps
and a velocity resolution of 0.65 \kmps respectively.

\nocite{stanford90}
\nocite{Wilson2000}

Our map is centered at the overlap region at R.A.= $12^h01^m54.8^s$, DEC=
$-18^o52'55''$ (J2000) and includes the two nuclear
regions. We used a grid size of 8.5$''$ which is a quarter of the beam
separation of the 2 $\times$ 2 array. Sixteen telescope pointings, each with
10--20 minutes integration, were required
to provide a complete 64-point map with angular extent $68'' \times
68''$, though slightly under sampled.  Our final map comprises three such
64-point maps covering most, but not quite the entire CO emitting region of
the two galaxies. The spectrometer band was
centered at V$_{LSR}$ = 1560 \kmpsnospace. The data were taken by position switching
in azimuth using a beam throw of 5$'$. This beam throw is sufficient
to position the reference beam well off the source at all map
positions. The
antenna temperatures were converted to main beam temperatures ($T_{mb}$) using
a main beam efficiency $\eta_{mb} = 0.46$. As a standard procedure, telescope
pointing was checked every 1.5 hours by observing SiO maser lines from
nearby evolved stars with the S40 receiver at  43 GHz.  The pointing
precision was typically 3$''$(rms) at both azimuth and altitude as the wind speed was low during 
most of our observing time.

\nocite{Kutner81}
\nocite{Ulich76}

The data were reduced using the NRO software NEWSTAR and then output
to the Astronomical Image Processing Software (AIPS) package for
further analysis. Figure 1 shows the grid spectra of $ ^{12}$CO J=1--0.  
A linear baseline was subtracted from each spectrum and
the spectrum was smoothed to 10.45 \kmps (3.9 MHz). Most of
the spectra were taken during the first observing run with the best 
weather conditions (T$_{sys}$ = 350 -- 600 K). 
Those points in the
southwest (bottom-right) region were taken during the second run and
the noise level was higher. Some positions were observed  in
both runs for comparison. The profiles  generally agreed
with each other and these have been co-added to generate the final spectra.

\section{The JCMT Observations}


Most of the $ ^{12}$CO and \thCO J=3--2 and 2--1 data were taken using the James Clerk
Maxwell Telescope (JCMT) during two observing runs during 1998 March
27 -- 28 and December 20 -- 23. Additional data for the \thCO J=3--2 were obtained in 2000 January.
Receiver B3 was used to
observe the $ ^{12}$CO and \thCO J=3--2 transitions at 345.796 GHz
and 330.581 GHz respectively, while receiver A2 was used to observe the $^{12}$CO and \thCO
J=2--1 transitions at 230.538 GHz and 220.399 GHz respectively.  The B3 receiver has dual polarized
mixers, permitting an increase in sensitivity by a factor of $\sqrt{2}$
when observing in the dual-channel mode. The system temperatures measured were $\rm T_{\rm sys}\sim 500$ K 
for  receiver B3 and 
$\rm T_{\rm sys}\sim 400$ K for receiver A2.  The weather conditions were very good during 
the 1998 March run, and values
 as low as $\rm T_{\rm sys}\sim 280$ K were obtained
for receiver B3.
 The Digital Autocorrelation
Spectrometer (DAS) was employed which provides a total usable bandwidth
of 920 MHz (802\kmpsnospace) for B3 and 
760 MHz (980 \kmpsnospace)  for A2. The frequency resolution was 0.625~MHz, 
which corresponds to a velocity resolution of 
0.545 \kmps at 345 GHz and 0.84 \kmps at 230 GHz. 
 The beam shape of the telescope is closely
 approximated by a gaussian with HPBW=14$''$ at 345 GHz and 20$''$ at 230GHz.

All the data  were taken by
beam-switching in azimuth using a beam throw of 120$''$. 
Focus and pointing were monitored frequently by observing bright
 quasars (3C273 or 3C279) and the pointing (rms) error was
 found to be $\sim 2''$ for B3 at 345 GHz and $\sim 2.7''$  for A2 at 230 GHz at both azimuth and altitude.
 The spectral line
 intensity is calibrated in terms of \TA, \ie
 corrected for atmospheric transmission and losses associated
 with telescope inefficiencies and rearward scatter. Conversion to $T_{\rm mb}$
was done by using the formula $T_{\rm mb}
= T_A^*/\eta_{mb} $, where the main beam efficiency $\eta_{mb}$ is 0.62 for B3 and 0.69 
for A2 according to the JCMT users guide.

The $ ^{12}$CO J=3--2 map was 
based on the region mapped at CO J=1--0 with the NRO 45m. We first produced a 7
$\times$ 7 grid map of the overlap region centered at R.A.=
$12^h01^m54.8^s$, DEC= $-18^o52'55''$ (J2000), then mapped the nuclear
regions of NGC~4038 and NGC~4039 with a 5 $\times$ 5 grid and a 3 
$\times$ 3 grid, respectively.  
With a spacing of 7$''$, we have fully
sampled the majority of the emission in this system. 
The integration time per point is
typically 2--8 minutes depending on the signal.  The spectrometer
band was centered at $\rm V_{\rm LSR} = 1560$ {\rm\,km\,s$^{-1}$}. 
Measurements of the CO J=3--2 spectral line of
IRC+10216 at $\rm V_{\rm LSR}=0$ km s$ ^{-1}$ were used to estimate
the calibration uncertainty by comparison with standard spectra in the JCMT archives. This
uncertainty is $\sim$ 10\%. Finally, we repeatedly monitored the  spectral line emission of the
map center throughout our runs and found  excellent agreement among all
   the line profile measurements.

  The spectra were reduced using the JCMT software SPECX and
a linear
baseline was removed from each spectrum. 
Figure 2 presents the observed $^{12}$CO J=3--2 profiles with a grid interval of $7''$.  
To permit detailed comparison with
the $^{12}$CO J=1--0 map, the $^{12}$CO J=3--2 data cube 
was imported into AIPS, re-gridded to $8.5''$ sampling, and convolved to
$15''$ spatial resolution. The re-gridded cube of the $^{12}$CO J=3--2 data 
were then smoothed to a velocity resolution of 
10.86~km~s$ ^{-1}$ (12.50 MHz). The 
resulting spectra are shown superposed on the corresponding CO J=1--0 
spectra in Figure 3. Since the $^ {12}$CO J=3--2 map is fully sampled, 
the re-gridded  profiles in Figure 3 are similar to the original ones in Figure 2. 
Figure 4(a) shows the integrated intensities with CO J=1--0 contours 
overlaid on the J=3--2 maps in gray scale form, and Figure 4(b)
shows a contour and gray scale map of the ratio of integrated line 
intensities.


Two spectra of \thCO J=3--2 were obtained --- 
one at the center of the overlap region, which is also the center of the $ ^{12}$CO maps, 
and the other at the nucleus of NGC~4038.
 The  total
integration time for the overlap region and NGC~4038 nucleus 
was  1.5  hrs
and 1.0 hrs, respectively.
In order to obtain a reliable  estimate of the line
intensity ratio of $ ^{12}$CO/\thCO J=3--2,  
 we  also performed a measurement  of the $ ^{12}$CO  J=3--2 line emission
at the same positions. The differences between the $ ^{12}$CO J=3--2 spectra observed at different times 
were within the calibration
uncertainties observed for B3 at this frequency.
Figure 5
presents the $ ^{13}$CO J=3--2 spectra
together with the  re-observed $ ^{12}$CO J=3--2 spectra
from the corresponding positions.


A 5 $\times$ 5 grid map with 10$''$ spacing at the
 overlap region centered at R.A.=
$12^h01^m54.8^s$, DEC= $-18^o52'55''$ (J2000) and one point each at the two
nuclear regions were
 observed at the $ ^{12}$CO  J=2--1 transition.  The integration time
 per point was 2--5 mins. \thCO  J=2--1 data were taken
 at the two nuclei and at three selected points 
in the overlap region with offsets 
(0, 0), (0, 10$''$) and (0, $-10''$) relative 
to $ ^{12}$CO  J=2--1  map center.
The $V_{LSR}$ was 1560 \kmps and the integration time was 40 minutes 
for each point.
Figure 6  presents the $ ^{13}$CO J=2--1 profiles
overlaid on the  $ ^{12}$CO J=2--1  profiles
at the positions where both measurements were made.
The $ ^{12}$CO J=2--1 profiles at other positions  can be found in Zhu (2001).

\section{Comparison between the $ ^{\bf 12}$CO J = 1--0 integrated intensities with previously 
published results}

Our CO map is concentrated mainly on the  nuclear regions of NGC~4038 and
NGC~4039,  and on  overlap region centered at 
R.A.= $12^h01^m54.8^s$, DEC= $-18^o52'55''$ (J2000). It covers most of the
emission from the galaxies, but its spatial coverage is not complete. 
For example, it does not extend to the western 
spiral arm of NGC~4038, and thus does not cover the CO emission associated
with the blue arc of star formation in this region.
\nocite{Strong88}
\nocite{Aalto95}
The total CO J=1--0 flux in our map is 2370 $\pm 310$ Jy \kmpsnospace, about 5 
times greater
than that (430 Jy \kmpsnospace) detected by Stanford \etal  (1990), and also
2 times greater than the total flux reported by Wilson \etal (2000). This 
value is in excellent
agreement with the single
dish measurements covering the two nuclei and the overlap region conducted by us using 
the NRAO 12m telescope (Zhu 2001), by 
Aalto \etal (1995) using the 15m Swedish European
Submillimeter Telescope (SEST)  (FWHM $\sim 45''$), and by Young
\etal (1995) using the Five College Radio Astronomy Observatory
(FCRAO) 14m telescope (FWHM $\sim 50''$). Based on their more complete
but much lower resolution (FWHM $\sim 55''$) map, covering a $200'' \times 150''$ region, Gao \etal  (2001) 
report a total CO flux of 
3595 $\pm$ 195 Jy \kmpsnospace, about 50\% higher than
in our map.  We have compared their measurements with ours
for the region in common and found good agreement between the two
data sets. To get an upper limit on the flux of CO emission possibly missed
in our map, we assume that the unobserved  emission within a $200'' \times 150''$ region 
has a low brightness temperature equivalent to twice the noise 
level seen in the 45m map, which is $\sigma$ = 5 K km s$^{-1}$. The conclusion
is that the missed flux under these circumstances would be about 2300 Jy km
s$^{-1}$. This estimate is indeed comparable to the difference between our
measured value for the integrated flux and that reported by Gao \etal (2001).     

Adopting a standard CO-to-H$_2$
conversion factor $X$ of 2.8$\times 10^{20}$~H$_2$~cm$^{-2}$ (K \kmpsnospace) $^{-1}$,
or M(H$_2$)=4.78$\times (L_{\rm CO}$/K~\kms~pc$^2)~\ms$ (Bloemen \etal  1986; Strong et
al. 1988), the integrated intensity in our map corresponds to a molecular 
gas mass (1.1 $\pm 0.14 $)$\times 10^{10}~\ms$.  The overlap region 
( $35'' \times
50''$) contains 910 $\pm 59 $ Jy \kmps or 42\% of the total CO flux,
indicating $\sim 4.2 \times 10^{9}~\ms$ of molecular gas in this $3.4
\times 4.9$ kpc region, which is 1.9 times more than that in the
OVRO map by Wilson \etal (2000) in the same region. The two nuclei of
NGC~4038 and NGC~4039 contain $\sim 1.8  \times 10^{9}~\ms$ and $\sim
9.5 \times 10^{8}~\ms$, respectively.

    \leavevmode


\section{Kinematics}

Our CO maps at both J=1--0 and J=3--2 transitions cover mainly the
overlap region and the two nuclear regions.  The goals of our
kinematic study are to distinguish different overlapping gas components
and gain insight into the origin of the large amounts of molecular gas in
the overlap region. A complete study of the velocity field of the
entire interacting system would require a kinematic model, and 
is beyond the scope of this paper.

From the channel maps of the $^{12}$CO J=1--0 and $^{12}$CO J=3--2 data in 
Figures 7(a) and 7(b), 
we can identify three components in the overlap region corresponding to super
giant molecular complexes (SGMCs) in the OVRO maps by Wilson \etal (2000).  The 
first component C1 (indicated by a star),
centered at (+5$''$, +5$''$), occupies the velocity range V = 1350 -- 1630 \kmpsnospace. The second
component C2 (indicated by an empty triangle) is at ($0''$, $-10''$) and
 V = 1410 -- 1650 \kmpsnospace. The third component C3 (indicated by a cross) is
at ($-5''$, +25$''$) and V = 1410 -- 1630 \kmpsnospace. C1 and C2 appear to be associated with the disk
of NGC~4039, whereas C3 appears to be associated with the region of overlap between the two
disks.  C1 and C2 are equally strong, while C3 is weaker. All three
components show very little variation in position over a velocity range of
240 \kmpsnospace. Closer examination of the line profiles (Figures 1 and 2)
reveals a double-peaked feature in C1, so this component can be
further divided into two sub-components. These are spatially
resolved into two SGMCs in the OVRO
interferometer map which are named as SGMC1 and SGMC2 by Wilson \etal
(2000).  The component C2 is also resolved into three SGMCs, named
SGMC3, SGMC4 and SGMC5 in the OVRO map. The molecular gas in the
component C3 is more diffuse than C1 and C2, and there are no named
SGMCs associated with it, though it is also detected in the OVRO
map.

The line profiles associated with C3 are striking, especially in the region near the
line $\Delta \delta = 20''$, or DEC(J2000) = $-18^o52'35''$.
This line is coincident with the southern edge of the NGC~4038
disk where it begins to overlap with NGC~4039, and 
hereafter we denote this line 
as ``OV-line''.
From the overlaid  $ ^{12}$CO J=3--2
and 1--0 line profiles in Figure 3, it can be seen that 
the profiles for the two transitions  generally agree quite well with
each other in the central and southern part of the overlap region with
declination offset $\Delta \delta < 10''$ from the map center. However,
in the region near the OV-line, the
J=3--2 transition peaks are offset from the J=1--0 peaks, implying at
least two velocity components with different excitation.  More
remarkably, the stronger CO J=3--2 peak switches from the high velocity  to
the low velocity feature when
crossing the OV-line from south to north, 
while the J=1--0 profiles do not
show such a dramatic change. This indicates a sharp excitation
gradient for both the high and low velocity components near the OV-line.
As mentioned above, this line appears to be the 
border between the two optical disks.   Therefore the degree of excitation in the two disks may be
very different from one another.

The velocities in the overlap region are all lower
(blueshifted) than the velocity of the two nuclei, both of which have a
similar velocity V = 1640 \kmpsnospace. The almost identical velocity for the
two nuclei suggests that the relative motion between these two
galaxies is in the plane perpendicular to the light of sight, or that the
two galaxies are at the point of turning around from receding to
approaching each other.  This pattern is consistent with that indicated by 
the
measurements of HI (van der Hulst 1979; Hibbard \etal 2001) and
H$\alpha$ (Amram \etal 1992). 
\nocite{Amram1992}
\nocite{Hulst79}
From the 
kinematics of the tails and HI velocity field of the spiral arms, Hibbard \etal
(2001) concluded that NGC~4039 is in front of NGC~4038 (closer to the
observer) and both galaxies are rotating anti-clockwise.  Higher
spatial resolution data from CO J=1--0 (Stanford \etal 1990) and
H$\alpha$ line emission (Amram \etal 1992) show a velocity gradient
around the nucleus of NGC~4039, which suggests that the differential
rotation is redshifted  southwest and blueshifted northeast of
the nucleus.  This rotation pattern can partly account for the blueshifted
velocity we see in the overlap region, but cannot explain
the complicated spectral profile in Figure 3, especially for the
component C3.  The blueshifted velocity component in C3 extends
continuously from the NGC~4038 nucleus down to the disk of NGC~4039. This
could be material being accreted onto NGC~4039 from the bridge
connecting the two disks. Similar gas components with the same velocity
range in the same region are also seen in the VLA HI image (Hibbard
\etal 2001).  In addition, the overlap region is also the place where
the northern tail of NGC~4039 connects to the inner disk. The
infalling or accreting material can cause strong shocks and greatly
distort the velocity field in this region. Further n-body
simulations of the encounter including gas hydrodynamics with high
spatial resolution are necessary to explain the
kinematics in the overlap region.

\section {Ratios of the integrated line intensities}

We define the ratios of integrated main-beam temperature as follows:

$$r_{21} = \int T_{\rm mb} (^{12}CO  J=2-1) dV / \int T_{\rm mb} (^{12}CO J=1-0)dV $$
$$r_{31} = \int T_{\rm mb} (^{12}CO  J=3-2) dV / \int T_{\rm mb} (^{12}CO J=1-0)dV $$
$$r_{32} = \int T_{\rm mb} (^{12}CO  J=3-2) dV / \int T_{\rm mb} (^{12}CO J=2-1)dV $$
$$^{13}r_{32} = \int T_{\rm mb} (^{13}CO  J=3-2) dV/\int T_{\rm mb}(^{13}CO J=2-1)dV
$$
$$R_{10} = \int T_{\rm mb} (^{12}CO  J=1-0) dV / \int T_{\rm mb} (^{13}CO J=1-0)dV $$
$$R_{21} = \int T_{\rm mb} (^{12}CO  J=2-1) dV / \int T_{\rm mb} (^{13}CO J=2-1)dV $$
$$R_{32} = \int T_{\rm mb} (^{12}CO  J=3-2) dV / \int T_{\rm mb} (^{13}CO J=3-2)dV $$

\subsection{ A gradient in the CO J=3--2/J=1--0 ratio}

After convolving the CO J=3--2 data from 14$''$ to 15$''$ to match the resolution of the 
CO J=1--0 data we can obtain reliable
line intensity ratios of $ ^{12}$CO J=3--2/J=1--0 ($r_{31}$). 
We used only those points with
good S/N ratio for both the $ ^{12}$CO J=1--0 and $ ^{12}$CO J=3--2
spectra.  Table 1 lists the positions, integrated intensities at CO J=1--0,
3--2 and the $r_{31}$  ratios, together with the uncertainties in the latter quantity.  
Figures 4(a) and 4(b) show
that the CO J=3--2 peak is offset from the CO J=1--0 peak  by $\sim 3''$ 
in different
directions in the two nuclei (to the north for NGC~4038 and south for
NGC~4039), which may possibly be due to a pointing error.  This would
result in an overestimate of the $r_{31}$ ratio north of NGC~4038 and south of
NGC~4039 nucleus by  approximately 10\%. Thus the high
 $r_{31}$ ratio in these regions should be interpreted with caution. 
In the overlap region the two map peaks appear coincident, indicating
that pointing error should be less than 3$''$.  We also convolved the $
^{12}$CO J=3--2 map to the angular resolution of $ ^{12}$CO J=2--1
(20$''$) to obtain the $r_{32}$ ratio, which should be less sensitive to
pointing errors. The distribution of the $r_{32}$ ratio is
similar to that of the $r_{31}$, so the gradient in $r_{31}$ seen in the
overlap region is most likely real.

The $r_{31}$ ratios in the nuclei of NGC~4038 and NGC~4039 are respectively $0.96
\pm 0.10$ and $0.77 \pm 0.10$, which is higher than the average value
$0.64 \pm 0.10$ in the nuclei of nearby spirals and starburst galaxies
(Devereux \etal  1994; Mauersberger \etal 2000). \nocite{Devereux94}
 The gradient in $r_{31}$ across the
overlap region is striking.  It is as low as 0.3 -- 0.4 in the northern
part of the overlap, but changes to $0.8 - 0.9 $ toward the southern edge
of the overlap region. The $r_{31}$ ratio also drops by a factor of
two toward the east and west edge of the overlap region. Figures 8(a) and 8(b) show respectively 
the $r_{31}$ contours overlaid on the K-band NIR image (Hibbard \etal
2001) and MIR 15 \um image from ISOCAM (Mirabel \etal 1998). The NIR image 
reveals two young clusters in the southern overlap region which are $\sim 4$ 
times brighter than the clusters in the northern overlap. The bright knot
in the southern overlap in the 
ISO MIR image indicates large amounts of hot dust
which is likely to be heated by the young clusters there. 
The derived $r_{31}$ ratios are spatially 
correlated with the
strong NIR and MIR emission, suggesting that high excitation regions 
are closely associated with the warm gas/dust.

As mentioned in \S 5, there is also an excitation gradient 
associated with different velocity 
components near the OV-line  DEC(J2000) = $-18^o52'35''$ (see Figure 3).
North of this line, the low
velocity component has a higher $r_{31}$ ratio than the high velocity
ones. This pattern is also seen in the disk of NGC~4038, east of the nucleus.
On the other hand, south of the OV-line, including the
disk and nucleus of NGC~4039, the $r_{31}$ ratio is similar for
both the low and high velocity components.
The most striking profiles are in the vicinity of the OV-line,
where the low velocity component is undetectable in CO J=3--2  ($r_{31} < 0.1$), whereas  
$r_{31}$ is as high as $0.8-0.9$ in the
high velocity component. Since CO J=3--2 is generally associated with
warmer and denser regions, the high velocity component (with high
$r_{31}$ ratio) in this region is most likely to originate in molecular clouds in the NGC~4039 disk, while the low velocity component could originate in
 the halo where the temperature and density may be too low
to generate significant CO J=3--2 emission.
 Considering that the low velocity component 
corresponds to a high
blueshifted velocity, and the fact that near the OV-line the two 
optical disks appear to be overlapping, 
this component could also be the gas in the bridge 
being accreted by NGC~4039.   Detailed
modeling including both gas dynamics and radiative transfer is needed
to test this hypothesis and explain the excitation gradient over the
velocity range.

\nocite{Aalto94}

\subsection{ $^{12}$CO/\thCO line intensity ratios}

At some selected points in the overlap region and the two nuclei, isotope emission \thCO J=3--2
and J=2--1 was detected making it possible to estimate the $R_{32}$ and $R_{21}$. These
ratios should be relatively free from errors in pointing and
absolute flux calibration since the two lines were observed immediately
one after the other with the same receiver. Thus the primary contributors
to the uncertainty are thermal noise and baseline errors. The derived $R_{32}$
ratio corresponds to 14$''$ angular resolution. If this ratio holds also for the
20$''$ resolution associated with $R_{21}$, and assuming that the \thCO J=3--2 emission follows the
distribution of $ ^{12}$CO J=3--2 emission, 
we can estimate the \thCO J=3--2 line intensity
for a 20$''$ beamsize and derive the \thCO J=3--2/ J=2--1 ratio, which is
$0.77 \pm 0.23$ for the center of the overlap region and $0.60 \pm 0.20$ for the nucleus of NGC~4038.
Table 2 lists the positions, line ratios and   their   associated
uncertainties  for the five points with \thCO line
measurements. The higher resolution $^{12}$CO J=3--2 emission has been convolved
to 20$''$ to match the measurements at lower resolution. 
The convolved $r_{31}$ ratio is slightly lower than that 
in Table 1 which corresponds to 15$''$ resolution.

The largest value of the ratio $R_{21}$ = 29 is at
offset (0, +$10''$) which is in the overlap region, 34$''$ (3.3 kpc) away from the
nucleus of NGC~4039. This value is a factor of 2 higher than the mean value of
 $R_{21}$
for normal spiral galaxies and starburst galaxies (Aalto \etal 1995).
The value of $R_{32}$ at the overlap region (0,0) central position is also
higher by a factor of 3 than those found by Wall \etal
(1993) in the nuclear regions of nearby starburst galaxies.  To our knowledge,
this is the
first published case in which  high $R_{21}$ and $R_{32}$ ratios have ever
been found outside the nucleus of a galaxy.

It is well known that interacting galaxies, especially mergers, tend
to have higher $R_{10}$  integrated line
intensity ratios than normal spiral galaxies (Aalto \etal  1991, 1995,
1997; Casoli \etal  1991, 1992a, 1992b; Hurt
\& Turner 1991; Turner \& Hurt 1992; Garay, Mardones, \& Mirabel 1993;
Henkel \& Mauersberger 1993; Henkel \etal  1998). 
\nocite{Casoli92a}
\nocite{Casoli92b}
\nocite{Casoli91}
\nocite{Hurt91}
\nocite{Turner92}
\nocite{Garay93}
\nocite{Henkel93}
\nocite{Henkel1998}
\nocite{Taniguchi98}
\nocite{Taniguchi1999}
The origin of this so-called ``missing $^{13}$CO'' problem (Casoli \etal 1992b) is not yet
well understood. Aalto \etal (1991, 1995, 1997) 
explained this effect by a relatively low
optical depth at the $^{12}$CO J=1--0 transition, but many
investigators attribute it to the  high $[\rm ^{12}CO/
^{13}CO]$ abundance ratio (\eg Henkel \etal 1998; Taniguchi \& Ohyama 
1998, Taniguchi, Ohyama \& Sanders 1999). Such abundances can be produced by (1) the inflow of
unprocessed disk gas with high $ ^{12}$CO /\thCO abundance ratios, (2) the
$^{12}C$ enhancement caused by selective nucleosynthesis in massive
stars (\eg Henkel \& Mauersberger 1993), (3) selective
photodissociation of the $ ^{13}$CO molecules in the UV-intense
starburst environment (\eg Fuente \etal   1993).

\nocite{Fuente93}

From starburst chemical evolution models Henkel~\&~Mauersberger (1993) 
predict that large
amounts of low $^{13}$C abundance material should exist in the outer disks of
spirals. However, the extreme weakness of the \thCO J=1--0 line makes it
difficult to detect in the outer region of normal spirals.  This
problem is alleviated by galaxy interaction which can enhance the CO
emission on the outer disk through cloud collisions and induced
starbursts.  Our finding of high $R_{21}$ and $R_{32}$ ratios in the overlap
region of the Antennae is consistent with the gas
infall and chemical evolution scenarios. The region with the highest values of $R_{21}$ is associated with
the youngest starburst clusters in the HST image, whose ages are
estimated to be less than $5\times 10^6$ years (Whitmore \etal
1999). These scenarios can also naturally explain the gradient of the
$R_{21}$ ratios, \ie highest in the north and lowest in the southern part of the
overlap as well as in the two nuclei. The kinematic data (see \S 5)
suggest that most of the gas infall/accretion activity is taking
place in the northern part of the overlap region near the OV-line 
at  DEC(J2000) = $-18^o52'35''$.

However, we note that many mergers with a high $R_{10}$ ratio do not necessarily have
a high $R_{21}$ ratio compared to other starburst galaxies (Aalto
\etal. 1995). This leads Aalto \etal (1997) to argue that the high
$R_{10}$ ratio is associated with small, warm but compact molecular
clouds with low-to-moderate optical depth ($\tau \sim $ 1) in the
$^{12}$CO J=1--0 line. With the measurement of both $R_{21}$ 
and $R_{32}$, we
can perform an excitation analysis to better constrain the
$ ^{12}$CO/\thCO abundance ratio.

\section{Excitation analysis}

 The variation of the $r_{31}$, $R_{21}$, and $R_{32}$ ratios in 
the Antennae system
 indicates dramatic changes in excitation conditions. These ratios
 cannot readily be explained under the assumption of LTE.  To estimate
 the physical parameters of the molecular gas, we have
 employed a large velocity gradient (LVG) model (\eg Goldreich \&
 Kwan 1974). In such models it is assumed that the systematic
 motions rather than the local thermal velocities dominate the
 observed linewidths of the molecular clouds.  
We use
 this model to fit the observed line ratios in Table 2, for different
 combinations of (\Tk, n(H$_2$), $\Lambda $), where \Tk  is the kinetic
 temperature and $\rm \Lambda =X_{CO}/(dV/dR)$, with X$_{CO}$=[$
 ^{12}$CO/H$_2$] being the fractional abundance of $^{12}$CO with respect to
 H$_2$ and dV/dR the velocity gradient. The optimum set of parameters
 is determined by  minimizing $\chi ^2$.

An extensive grid of LVG models was searched with a parameter range of
$\rm T_{\rm kin} =5-200$ K ($\Delta\rm T_{\rm kin}=2$~K), $\rm
n(H_2)=(0.1-10^4)\times 10^3$~cm$ ^{-3}$ ($\rm \Delta log n(H_2)=0.5$),
$\Lambda = (0.1-10^2)\times 10^{-6}$ \XdVunit
($\Delta \rm log\Lambda = 0.5$), and $\eta$ = \1213CO = 20 -- 100 
($\Delta \eta = 5$). These
ranges cover all possible conditions found in the GMCs of our
Galaxy as well as in external galaxies.  The results are summarized
in Table 3. Columns 2--4 contain the LVG parameters noted above, column 5
contains the CO column density, column 6 the optical depth at the J=1--0
transition, column 7 contains the brightness dilution factor (defined as the ratio
of the observed main beam brightness temperature to the radiation temperature
given by the LVG model), column 8 contains the assumed isotope abundance ratio, and 
column 9 shows the value of $ \chi $ associated with the fit.

\subsection{Single component modeling}

The center position of the overlap region (offset (0,0)) and the
nucleus of NGC~4038 have both \thCO J=3--2 and J=2--1 measurements, which
enable us to better constrain the physical parameters. At the offsets
(0, +10$''$) and (0, $-$10$''$) in the overlap region and the nucleus of NGC
4039 where only $R_{21}$ was measured, we assume that the $R_{32}$ ratios at
these positions are similar to that of the (0,0) position, but with an
uncertainty of 50\%.  Table 3 lists the estimated physical parameters
for these regions using a single component model.

Table 3 shows that a
 single component model can fit the
line ratios in the NGC~4038 nucleus reasonably well, with $\chi_{min} $= 1.2.
The best fit corresponds to
\Tk= 43 K, \nh2 = $  3.5 \times 10 ^3$ \cm3, 
 $ \Lambda =  3.2 \times 10^{-6} $\XdVunit  
 which gives a
$r_{21}$ = 1.1 ,
$r_{31}$ = 0.82, 
$R_{21}$ = 13.7,
$R_{32}$ = 20  and 
$^{13}r_{32}$ = 0.51.
However, for the overlap region where the $R_{21}$ and $R_{32}$
ratios are very high,  no acceptable  fit
can be found with a single component model. 
 The closest fit ($\chi_{min} = 2.2$) 
corresponds to moderate optical depth gas 
with $\tau
\sim $ 0.9, 1.5, 3.4 at the $^{12}$CO J=1--0, J=2--1, and J=3--2 transitions respectively (Table
3), with an abundance ratio $\eta = 70$.  This fit yields $^{13}r_{32}$
= 0.48, considerably lower than the observed value of 0.77. Thus unless
we have overestimated $^{13}r_{32}$ by 40\%, a single component model cannot
reproduce the observed line intensity ratios.  In general the 
ranges $\eta$ = 40 -- 200, \Tk = 28 -- 160 K, \nh2 = $(1 - 6 ) \times 10^3 $ \cm3 
and X$_{CO}$/(dV/dR) = (1 -- 10) $ \times 10^{-6}$ \XdVunit 
can fit the
ratios equally well with $\chi < 3$. 
When  $\eta$  is less than 40, no fit is possible.
This indicates that the probability for a larger $\eta$ is high 
in the overlap region.

\nocite{Wilson94}
\nocite{Henkel94}
\nocite{Henkel93b}

\nocite{Irwin92}
\nocite{Eckart90}
\nocite{Peter98}
\nocite{Mao2000}

Further to this point, it is probable that two components are necessary to model the overlap
region.
A lower limit on gas density can be set by the $^{13}r_{32}$ ratio.  If the
\thCO emission is from the same component as  $^{12}$CO, the \thCO J=3--2
must be optically thin (since $R_{32}$ $>> 1$) and normally excited by
collisions.  The critical density of \thCO J=3--2 is relatively high
($\sim 4 \times 10^4$ cm$^{-3}$) and moderate gas density is needed to
excite enough \thCO \ molecules to the J=3 level.  
The result is that \nh2 must be greater than $1 \times 10^4$ \cm3 to
satisfy $^{13}r_{32}$ = 0.8.
However, at this
density, the optically thick line $ ^{12}$CO J=3--2 is nearly
thermalized due to radiative trapping and the $r_{31}$ ratio should be close to
unity.  
The result is that with a $r_{31}$ ratio of $0.78 \pm 0.08$, 
the
$^{12}$CO emission is most likely from a gas density lower than $10^4$
\cm3.  This situation is also true for other kinetic temperatures we
searched from 10 K -- 200 K.

 Such discrepancies clearly demonstrate that a single component model
fails to describe the complicated physical conditions in the ISM of
the Antennae. This is not surprising since the optical depths of $^{12}$CO
and \thCO are significantly different and they may be probing different
regimes of the molecular cloud with different physical conditions. In addition, the complex 
line profiles in the overlap region indicate the presence of more than one component.

\subsection{A two-component model for the overlap region}

Our analysis in the previous section suggests the presence of at least two
components: a low density component [\nh2 $ < 10^4$ \cm3] dominating
the $ ^{12}$CO  emission and a high density component [\nh2 $ > 10^4$ \cm3]
dominating the \thCO emission.  
This is consistent with the findings
of other investigators from the studies of other galaxies (\eg Aalto
\etal  1994, 1995;  Wall \etal 1993). \nocite{Wall93}  As suggested by Aalto
\etal (1995), in regions with strong tidal force, such as the overlap
region of the Antennae and the galactic nucleus with strong
differential rotation, gas of low density will not survive as clouds, but
will be torn apart and sheared into filaments. Dense gas, on the other
hand, will remain in cloudy structures. Alternatively, the two-phase
model can also be a combination of photon-dominated regions (PDRs)
(\eg Tielens \& Hollenbach 1985) on the surface of $^{13}$CO-emitting
dense cores. The low density phase is more diffuse and has a much
larger filling factor, which makes it the dominant contributor 
to the $^{12}$CO
emission. The dense cores have a much smaller filling factor, but
may have significant optical depth (i.e., $\tau \sim 1$) for $
^{13}$CO, thus contributing most to the observed $ ^{13}$CO emission.
Evidence for such dense molecular gas has been uncovered in a number
of studies where \thCO is found to be optically thick (\eg White
1997).

\nocite{White97}
\nocite{Tielens85}

A two-component model will readily fit the observed line ratios. 
These two components are not resolved either in velocity or in space, and 
we assume the same values of
$\eta$ and $\Lambda$ for both 
components. Taking the low density component A as the one that best
fits the data in the single component model (Table 3), we find that
a component B with \nh2= $1\times 10^5$ \cm3 and \Tk = 90 K can fit
the observed line ratios within the measurement errors.  This
component has a $^{13}r_{32}$ ratio of 1.7. The combination of
these two components with dilution factors (defined
as the ratio of the observed to model brightness 
temperatures) $f_A$ = $5.1\times 10^{-2}$ and $f_B$ =
$3.3\times 10^{-4}$ gives $r_{21}$ = 1.06, $r_{31}$=0.80, $R_{21}$ = 23,
$R_{32}$= 28, $^{13}r_{32}$ = 0.6.  In Table 4 we list four
combinations that can fit the observed data reasonably well.

The low density component generates most of the $^{12}$CO
emission, with ratios  $R_{21}> 26$ and $R_{32}  > 25$. The $r_{32}$
ratio is 0.7--0.8 in this component, but $^{13}r_{32}$ is smaller
than 0.5. Thus the high density component requires $^{13}r_{32}$ $>$ 1,
which implies warm
and optically thin \thCO emission.  It should have a filling factor
much smaller than that of the low density component, so that it has
little effect on the $^{12}$CO emission.
The small dilution factor of the dense component is consistent with the
weak HCN J=1--0 emission in the overlap region and the nuclei of NGC~4038
(Gao \etal 2001). Gao \etal (2001) find that the $L_{HCN}/L_{CO}$ is lower for the
Antennae compared with other starburst galaxies, implying that the fraction of
high density gas in this system is smaller than other IR luminous
mergers.

It must be noted that, by introducing a two-component model, there are
twice as many free parameters and thus too few observed line intensity
ratios to constrain them.  The data can be satisfied with  very
different parameters T = 30 -- 130 K,  \nh2$ = (1 - 8) \times 10^3$ \cm3
for the low density component and T=40 -- 200 K, \nh2 $ > 3 \times
10^4$ \cm3 for the high density one, with the assumption that both 
components have the  same  $\Lambda = 10^{-6} - 10^{-5}$ \XdVunit and
isotope ratios = 40 -- 100. The LVG fit of the line intensity ratios is useful
only as a rough guide on the average conditions expected in the region
covered by the telescope beam ($\sim 2$ kpc here), but it shows that 
the molecular gas that produces most of the $^{12}$CO emission
must have a low optical depth, whereas the \thCO emission probably originates in 
a different component with a much smaller filling factor. In addition, the
abundance ratio $\eta$ is high compared to Galactic molecular clouds.

Although the single-component model can fit the nuclear region of
NGC~4038, the results do not rule out the presence of multiple components here as well.

\subsection{Inferences from the $r_{31}$ ratio alone}

For most of the region mapped there are measurements of $r_{21}$ and $r_{31}$ only (\ie no
isotope measurements). However,  the $r_{21}$ ratio is 
close to unity indicating that  the
$^{12}$CO J=2--1 emission is thermalized. Thus it cannot
 provide much of a constraint on the physical parameters. However the
$r_{31}$ ratios are under unity for all regions, and are sensitive 
to excitation conditions. In the nuclear and southern overlap regions,
 $r_{31} \sim  0.8 $ and this value  
 sets a lower limit to the kinetic temperature
\Tk$ > 20 K $ since no LVG solution is possible otherwise. In the northern part of the overlap
region where the
$r_{31}$ ratio is found to be $\sim 0.4$, the lower limit for the gas
temperature is $\sim $ 10 K. When the kinetic temperature is greater than
the lower limit, 
the $r_{31}$ ratio also depends on the gas density.  
 For \Tk
=20--80K, a value of $r_{31}$=0.8 indicates $n(H_2) \approx 10^3$ \cm3,
while $r_{31}$=0.4 implies $n(H_2) \lta  10^3$ \cm3. 
Near the OV-line the low
velocity component has $ r_{31} < 0.1 $, this extremely low $r_{31}$ ratio 
suggests that either the molecular temperature is very low ( $< 10 K$ ) or
the gas density is very low ( $ \lta 5 \times 10^2$ \cm3).

\subsection{Molecular gas mass }

In \S 4 we estimated the amount of molecular gas based on the
assumption of the conventional Galactic CO-to-H$_2$ conversion factor.  
Now we can use the parameters derived from the LVG model to get a more
knowledgeable estimate of the molecular gas mass.

From Table 3, the total CO column density can be constrained to a
narrow range \eg $ N(CO)/\Delta V =$ \nh2( $\frac{X_{co}}{dV/dR}) f \approx
(1.5 - 1.9) \times 10^{15} $ \cmsq/\kmps for the combination of two
components in the central 20$''$ of the overlap region.  Integrating
over the 20$''$ area, we have \Mco $= (1.4 - 1.7) \times 10^8 (\frac
{\Delta V}{200 \, km\,s^{-1}})( \frac {10^{-4}} {X_{CO}}) \ms$, where 
$X_{CO}$ is
the abundance ratio. In the same region, use of the Galactic
conversion factor yielded $1.8 \times 10^9 \ms$. Similarly
for the nuclear region (20$''$) of  NGC~4038, we have \Mco $= 8.5\times 10^7 (\frac
{\Delta V}{100 \, km \,s^{-1}}) (\frac {10^{-4}} {X_{CO}}) \ms$. These 
values depend upon the unknown value for $X_{CO}$, which is generally 
assumed to be in the range $10^{-5}$--$10^{-4}$ for starburst galaxies (\eg
Mao \etal 2000, Booth \& Aalto 1998). At the lower end of this range, the
molecular masses are comparable to those derived from the conventional $X$
factor, but $X_{CO} = 10^{-4}$ yields molecular masses which are a factor
of 11--13 times lower. 

Unfortunately, few if any reliable measurements of
the abundance ratio exist for starburst galaxies. Reliance must be 
placed on observations and theoretical models of Galactic GMC's and star 
forming
regions (e.g. OMC-1 and Sgr B2). Typical values of $X_{CO}$ for such regions are near $10^{-4}$,
and range from $5 \times 10^{-5}$ to $2.7 \times 10^{-4}$ (\eg Blake 
\etal 1987, Farquhar \etal 1997, Hartquist \etal 1998, van Dishoeck 1998, van Dishoeck \& Blake 1998).
Thus it appears unlikely that $X_{CO} \le 5 \times 10^{-5}$ in starburst 
regions unless the metallicity is abnormally low. If $ X_{CO} \sim 10^{-4}$, 
then the 
values of $\Lambda \sim 10^{-6} - 10^{-5}$ in Table 3 imply the existence of
a low optical depth, such that $dV/dR \sim 10 - 100 \, km \, 
s^{-1} \, pc^{-1}$ for the constituent clouds. Though such large velocity
gradients  cannot
be ruled out by observation of individual clouds in starburst systems because
of inadequate spatial resolution, they seem perhaps physically implausible
even for disrupting clouds. It is possible that a combination of lower 
$X_{CO}$
and large velocity dispersion are responsible for the low optical depths. It
is likely that such clouds are virially unstable, and that the large 
velocity dispersions in the clouds give rise to a higher CO radiative 
efficiency (and hence lower $X$ factor). If $X_{CO} \sim 10^{-4}$ then the 
corresponding CO-to-H$_2$ conversion factors $X$ would be (5.1 -- 6.4) $ \times 10^{19} (\frac {10^{-4}}
{X_{CO}})$ cm$^{-2} [K km s^{-1}]^{-1} $  
for the central overlap and $2.3 \times 10^{19} (\frac {10^{-4}}
{X_{CO}})$ cm$^{-2} [K km s^{-1}]^{-1} $   for the nucleus of NGC~4038. 
These values are 5-13 times
smaller than the conventional $X$ factor $2.8 \times 10^{20} $ 
cm$^{-2} [K km s^{-1}]^{-1} $  
(Strong \etal 1988). These are likely to be lower limits, and it seems 
plausible to adopt a value for $X$ by, say, a
factor of $5$ lower than the Galactic value overall, so that the inferred total mass of molecular gas
in the Antennae system would be reduced to $2 \times 10^{9} \ms$.

\section{Comparison with sub-mm continuum}

Sub-mm continuum data were obtained by Haas \etal (2000) using SCUBA on
 the JCMT.
The angular resolution is 15$''$ at 850 \um and 7$''$ at 450 \um.  The
former is essentially identical to that
for our NRO and JCMT line data. Our CO J=3--2 line data thus permit a reliable estimate to be
made of the contribution to the SCUBA 850 \um brightness by this line which is contained within
the SCUBA bandpass filter.
Although SCUBA measures primarily the continuum emission from dust,
this molecular line
contribution must be removed before we
can interpret the continuum emission. This SCUBA equivalent flux for a line 
in the SCUBA band at the radial velocity of the Antennae galaxies was estimated using the SCUBA filter bandpass profile and a beamsize of 15''. The conversion
factor used for 15$''$ SCUBA beam is 0.735 mJy beam$^{-1}$ (K km/s)$^{-1}$.
Figure 9(a)  shows the uncorrected SCUBA 850 \um contours overlaid on the CO J=3--2
greyscale map. It may be seen that the
SCUBA image closely follows the CO J=3--2 distribution, especially in the
region of the CO J=3--2 peak, \eg in the overlap region and in the two nuclei.
Conversion of the CO J=3--2 brightness to SCUBA equivalent flux and integration of the brightness contours in each map over their common area 
shows that line emission accounts for approximately 46\% of the integrated 
850 \um flux in the region shown in Figure 9(a). This is somewhat greater 
than the contamination factor of 30\% estimated by Haas \etal (2000) based on a best
estimate in the absence of actual CO J=3--2 data.
The most serious contamination occurs in
the southern gas complex of the overlap
region (K1 in the map of Haas \etal 2000), where about 55\% of the flux is from
CO J=3--2. This region is coincident with the high excitation region
seen in Figure 4(b) indicated by the $r_{31}$ ratio. 
 
Figure 9(b) shows the result after subtracting the contamination by CO J=3--2 
emission 
from each pixel of the original SCUBA image. This image shows the emission from the dust alone
overlaid on the 
CO J=3--2 greyscale map. 
The two 850 \um knots in the
 overlap region (identified by Haas \etal  as K1 and K2) are more widely and clearly
 separated, with the northern knot K2 being slightly stronger than
 the southern knot K1.  K1 suffers more contamination from the strong CO
 J=3--2 emission since it appears at the same strength as K2 in the
 uncorrected map.  The corrected 850 \um flux ratio between K1 and K2 ~
S(K1)/S(K2) is 0.78. This ratio is similar to that for the 20 cm radio
 continuum, but is different from the ratio seen in the MIR which is 2.6 (Mirabel
 \etal 1998).  The 450 \um flux from Haas \etal (2000) is also much higher in K1 than in K2
 (no detection in K2), suggesting that the dust is hotter in 
 K1. As mentioned above, the CO J=3--2/J=1--0 ratio is also higher in K1
 than in K2.

Another striking feature in the corrected 850 \um map is the lack of
 dust emission concentrated in the two nuclei. The cold dust emission seems to
be diffusely distributed in the disks of NGC~4038. In NGC~4039, we see
almost no cold dust at all at the NGC~4039 nucleus. Instead, more dust
is found along the spiral arm southwest of the nucleus. This dust
complex is detected in both 450 \um and 850 $\mu$m, and thus is most likely to
be real. However, it is only a 4$\sigma$ detection, and more sensitive
measurements are needed to confirm it.

Color plates 1 and 2 show the overlay of the
corrected SCUBA 850 \um contours on the HST full 
color and H$\alpha$ images from Whitmore \etal (1999). 
A dust lane in the overlap region is clearly seen in the HST full color image (Color plate 1). 
The SCUBA 850 \um emission  coincides with the dust lane, and  young clusters
are mainly seen at the edge of the dust lane. Some
clusters are heavily obscured by dust. These clusters appear to 
be faint and red in the color image, but are clearly seen in the
 H$\alpha$ image in Color Plate 2. 
Since H$\alpha$ emission requires the
presence of O and B stars to ionize the gas and these stars last less
than 10 Myr, the presence of H$\alpha$ emission alone
guarantees that the region contains clusters younger than 10 Myr. The
size of the H$\alpha$ bubble can provide further information on the
age of the clusters. The H$\alpha$ bubbles within the 850 \um peak K1
and K2 are very compact, which suggest that this region is relatively
young, while the cluster complexes bordering K1 and K2 and those along
the northeast side of the dust lane have bigger bubbles and are probably
older. Apparently, there has not been enough time for the complexes to
blow away all the dust yet.
Whitmore \etal (1999) estimated that the age of these clusters is $\sim 5$  
Myr. 

In the H$\alpha$ image (Color plate 2), there appear to be more clusters near K2 than
K1, and yet K1 generates much more MIR and NIR emission than K2. Thus the clusters
in K2 may be so young that the star formation is still deeply
cocooned and hence almost invisible at optical or even NIR and MIR
wavelengths. The age of these clusters should be less than  5 Myr.

Figure 10 shows contours of the corrected SCUBA 850 \um (white) and 450 \um 
(black)
emission overlaid on the Nobeyama 45m CO J=1--0 greyscale map. The
differences between the dust emission and CO emission are remarkable.
In the overlap region, the CO J=1--0 peak is located between the two 850 \um
peaks, which is similar to the CO J=3--2 distribution. 
The 850 \um emission in the two
nuclei also seems to be offset from the CO J=1--0 peaks, especially for
the dust concentration southwest of the NGC~4039 nucleus where little
CO  J=1--0 emission is detected in the Nobeyama single dish or OVRO
interferometer maps by Wilson \etal (2000). However, we do see weak extended features in
the BIMA + NRAO 12m full synthesis CO J=1--0 map (Gruendl {\it et~al.}, in preparation)
in the position coincident with the dust emission. The difference
between the CO and dust 850 \um continuum suggests that the molecular
gas traced by CO is not closely associated with the cold dust
emission. This is consistent with our conclusion from the excitation
analysis in which the $^{12}$CO J=1--0 emission is mainly from a warm and
moderately opaque component.

Table 5 shows the total dust mass in the system computed 
using the corrected 850 \um flux.  The total flux is well fitted by a
one-component dust model with an emissivity law $\lambda^{-2}$ ($\beta=2.0$), 
dust temperature
T=28 K and mass $M_{dust} = 1.7 \times 10^7 \ms$.  
The new estimate of dust mass is only 20 percent of the value
suggested by Haas \etal (2000) ($ \sim 10^8 \ms$), which is attributable to
our larger estimate of the contamination by CO J=3--2 emission. Combining this with the gas mass of 
$2 \times 10^{9} \ms$, we obtain a gas to dust mass ratio of about $100$.

To fit the SED for the overlap region, mainly the two sub-mm knots, we
assume that the FIR fluxes at 60, 90, 120, 160 and 200 \um 
in this region are 35\%
of the total flux from Klass \etal  (1997). This fraction is estimated
by Haas \etal (2000) from the ISOPHOT 60 \um and 100 \um maps.  
A one-component model can fit the SED reasonably well (using 3$\sigma$
for K2 as an upper limit), 
with dust temperature T=28 K and mass $M_{dust} = 6.0 \times
10^6 \ms$.  This mass is a factor of 4.4 times less than 
that estimated by Haas \etal (2000). We have
also tried to fit  the SEDs of the individual knots  with the
assumption that FIR emission is  distributed equally between  K1 and K2. A
dust temperature of 28 K and $\beta$ =2.0 can fit K2 well within the
error range, but K1 is better fitted with a dust temperature of 35K,
$\beta = 1.0$  for $\lambda < 250 \mu$m and $\beta = 2.0$ for $\lambda > 250 \mu$m (Hildebrand 1983). 
The estimated dust mass is
listed in Table 5, with K2 containing slightly more dust mass than K1.

\nocite{Johnstone99}

\section{Origin of the molecular gas in the overlap region}

First of all, referring to the whole region between the two nuclei as
``overlap'' may be misleading. In \S 5 we showed that
the kinematic data at the CO J=1--0 and 3--2 transitions suggest that
the actual overlap occurs near the OV-line DEC(J2000) = $-18^o52'35''$.
Only the region near this line could have gas from both disks or 
in a bridge between two disks. The majority of CO and dust 
emission appears to be from
the off-nucleus gas concentration in the disk of NGC~4039. 
If the conventional $X$ factor is used, this concentration contains 
$\sim 3.2 \times 10^9 \ms$ of molecular gas, more than the total
\Mco in
the Milky Way. This region also 
contains more than 30\% of the total MIR, FIR, sub-mm and radio continuum
emission, which makes it outshine the two nuclei at all these
wavelengths.  This is the place that contains most of the effects
related to galaxy collision and the induced star formation. We refer to it as 
the ``southern overlap'' region to indicate its relation to the actual 
overlap region. 
  The origin
of the molecular gas in this region is crucial for understanding the
star formation activity and for predicting the further
evolution of this
on-going merger.
Our LVG analysis indicates that the molecular gas 
may be $\sim 5-10$ times smaller 
than that estimated using a conventional $X$ factor. In the following
we argue that the lower value provides a reasonable estimate of the molecular gas 
based on the probable dynamical history, the short lifetime of molecular cloud and
the small age constrained by the young clusters found in this region.

The dynamical models suggest that following the initial encounter, the
two galaxies separated and then later re-engaged to form the current
configuration (\eg Toomre \& Toomre 1972; Dubinski \etal 1996).  The
overlap region is not the gravitational center and the ``overlap'' is
just a projection phenomenon. Within the dynamical time scale ($10^7$
-- $10^8$ years), proper motion and rotation of the disks would cause a
relative displacement of the overlapping gas from the two galaxies. Hence it is difficult to understand how this
particular region outside the galactic nucleus could accumulate large
amounts of molecular gas. If this is an effect of the tidal force, we
should also see similar effects in the stellar or atomic gas
components in the same region, which is not the case.  On the other
hand, even if the accumulation of gas were possible dynamically,
such a large scale dynamical process in a disturbed interacting system
would unavoidably trigger massive starbursts in the GMCs which would
then quickly destroy the molecular clouds. Hence the accumulation
process of the molecular gas would not last longer than the life time
of molecular cloud, which is $\sim 10^7$ years.  Indeed, a number of
young clusters have been found in the overlap region and the age of
the clusters was estimated to be $\sim$ 5 Myr (Whitmore \etal 1999).
This is consistent with the time scale we estimated if those clusters
were formed during the process of molecular gas
formation/accumulation.  Taking the maximum sound speed of 10--100
\kmps for the molecular clouds in the galactic disk, the gas can move
only 0.1--1 kpc in $10^7$ years. Thus not much gas can be accumulated
in one region during such a short period of time.

Haas \etal (2000) explained the dust/gas in the overlap region with a 
``traffic jam'' model, which suggested that initial 
inelastic cloud collisions would build a kernel into which 
the other clouds of the rotating disks are running, 
and making huge amounts of gas  pile up together. 
This mechanism, however, can only build up to 10\% of the gas mass,
simply because the lifetime of the molecular clouds is
less than $10^7$ years which is only one-tenth of the typical rotational
 period for
spirals. Hibbard \etal (2001) suggested that the overlap region represents
material transferred from NGC~4038 to NGC~4039. Such a model can explain the 
blue-shifted velocity pattern in the northern overlap region. However, from
our CO J=1--0 and 3--2 kinematic data, the average blueshifted velocity is
150 \kmps and therefore the gas can move only about 1.5 kpc in 10 Myr. Therefore
only a modest fraction of gas in the overlap region
could have resulted from accretion from NGC~4038.

Consequently the SGMCs in the
southern or northern overlap region and the two dust knots seen 
in SCUBA 850 \um  most likely  formed independently 
out of material originating nearby.
 If the gas distribution
in the progenitor NGC~4039 is originally normal, the
vicinity of the southern overlap should not have more than one-tenth of
the total molecular gas of that galaxy, which is $\sim 10^9
\ms$.  The total HI gas in the Antennae galaxies is $6.3 \times 10^9
\ms$, but 64\% of this is associated with the tidal tails, and only
one-third of the remaining HI flux is from NGC~4039 (Hibbard \etal 2001). 
Hence the local atomic gas originally in the south overlap region is also limited. 
If a large amount of H$_2$ is converted from HI, it is difficult to understand how
to move a huge amount of atomic gas into the overlap region and convert it into H$_2$
in such an efficient way in a short time.

In summary, the conventional X factor results in an extraordinary large amount of molecular gas 
outside the galactic nucleus which is difficult to explain with the existing theories of 
gas dynamics and molecular chemistry. 
On the other hand, if the X factor is actually 5--10 times smaller, the 
molecular gas mass would be $\sim (3.2 - 16) \times 10^8 \ms$ which
can be easily accounted for by the aforementioned mechanisms. 
Hence the strong CO emission in the southern overlap region is probably
originated from the local molecular gas that 
is rendered more luminous by increased starburst activity.  
Our LVG
analysis points to the possibility that large velocity dispersion and large
filling factor of a diffuse gas component are associated with a low optical depth
and high CO luminosity.  The southern overlap is special only in that
it is adjacent to the overlap region.  Cloud collisions, shocks and
induced star formation in this region can all contribute to the
large velocity dispersion and create a diffuse component of molecular gas
 with excess CO J=1--0 emission.

\section{Summary}

We have obtained high quality, fully sampled, single dish maps of the
Antennae galaxies NGC~4038/9 
at the $^{12}$CO J=1--0 and 3--2 transitions with an angular
resolution of 15$''$ (1.5 kpc). This is so far the highest resolution single
dish data at both the J=1--0 and 3--2 transitions. $^{12}$CO J=2--1 data at the
two nuclei as well as in part of the overlap region with $20''$ angular
resolution have also been
obtained.  In addition, the \thCO J=2--1 and \thCO J=3--2 
line emission are detected at selected points in the two nuclei and 
 the overlap region. Our major findings are summarized below:

1. The integrated  $^{12}$CO J=1--0 flux in 
the Nobeyama 45m single dish
map is about twice that reported by Wilson \etal(2000) using the OVRO
interferometer.
The $^{12}$CO J=1--0, 2--1, 3--2 emission all peak in
an off-nucleus region adjacent to where the two disks overlap. This region
accounts for approximately 42\% of the total $^{12}$CO J=1--0 flux of the
inner disk of the Antennae. 

A conventional $X$ factor yields $\sim 4 \times 10^9 \ms$ for the mass of
molecular gas
in the overlap region. Such large amounts of molecular gas are
difficult to explain given the short lifetime of molecular clouds and
transient overlap time scale.

2. The ratio of the integrated intensities $^{12}$CO J=3--2/J=1--0 ($r_{31}$) varies across the
overlap region by a factor of 2.
The two nuclei and southern overlap region have a higher $r_{31}$ ratio,
which is coincident  with the
 strong MIR and SCUBA 450 \um emission in this region. 

3. The ratio $r_{31}$ changes from less than 0.1 to as high as
0.9
over the velocity range 
from 1350 \kmps to 1620 \kmps near the OV-line
 DEC(J2000) = $-18^o52'35''$ which is the
southern edge of the NGC~4038 disk overlapping with NGC~4039, indicating at least
two velocity components with completely different excitation in this region.
North of this line the excitation reverses between the high and low velocity 
components, \ie the low velocity component becomes the high excitation one.
The kinematics and excitation patterns suggest that the molecular gas north of
this line is from NGC~4038 and south of this line is from NGC~4039.

4. Both the $R_{21}$ and $R_{32}$ ratios are remarkably high in 
the overlap region. This is the first published case in which such high
$R_{21}$ and $R_{32}$ ratios are found outside a galactic nucleus.

5. LVG models indicate that $^{12}$CO and \thCO emission comes
from at least two different co-spatial 
components. The low density component has \nh2 $\sim 3
\times 10^3$ \cm3 and is responsible for most of the $^{12}$CO emission. This
component must have a low to moderate optical depth ($\tau_{10} \lta 1$), 
and/or a high \1213CO abundance ratio. 
The high density component has  \nh2 $\gta 1 \times 10^4$ 
\cm3 and contributes more to the \thCO emission.

6. Assuming a Galactic [CO/H$_2$] abundance ratio of $10^{-4}$, we find
 that the $X$ factor for converting CO-to-H$_2$
is $\sim 10$ times smaller than the 
conventional value in the overlap
 and  nuclear region of NGC~4038. Therefore, the molecular
gas mass could be overestimated by a factor of up to 10 in the overlap
 region and in the two nuclear regions if the conventional $X$ factor is 
used. However the mass of molecular gas derived by the LVG method used here is
inversely proportional to the assumed fractional abundance of CO.

7. The $^{12}$CO J=3--2 line emission contributes significantly to
the SCUBA 850 \um continuum flux. For the Antennae galaxies, the ratio
of the contribution from $^{12}$CO J=3--2 emission to the SCUBA 850 \um flux is $\sim$ 46\% on
average, and it varies by a factor of two across the system.  
After correcting for the $^{12}$CO J=3--2 contamination, 
the dust emission at
 850 \um detected by SCUBA is consistent with the thermal emission
 from a single warm dust component with $\beta=2$ and $M_{dust} = 1.7
 \times 10^7 \ms$. This value is a factor of more than six times lower than that
 estimated by Haas \etal (2000) using a lower estimate for the contamination by $^{12}$CO J=3--2 line emission.

8. The corrected 850 \um continuum emission is not tightly  correlated with
 the CO J=1--0 emission.
This is perhaps related to the conclusion that the CO emission comes predominantly from a warm
diffuse component which does not necessarily follow the distribution of the cold dust. The
enhanced CO luminosity in the overlap region may be caused by this warm component heated by
star formation.

The authors wish to thank the staffs of the JCMT, NRAO and NRO for 
their generous
assistance. We also thank Dr. B.C. Whitmore for providing the HST images,
Dr. J. Hibbard for providing a HI VLA map and a K-band optical image, 
Dr. C.D. Wilson for providing a CO J=1--0 map made
with the OVRO array, 
 and Dr. M. Haas for providing maps of the 450 and 850 \um 
continuum emission made with SCUBA at the JCMT. We also thank an anonymous
referee whose suggestions led to a significant improvement of the paper. This research was supported 
by a research grant to ERS from the Natural Sciences and Engineering Research
Council of Canada. 
 
\clearpage

\parindent 0pt

{\large \bf Figure Captions}

{\bf Fig 1}
{The grid map
of all the $ ^{12}$CO J=1--0 spectra. The temperature scale is $\rm
T_{\rm mb} = -0.2...1.0$ ~K, the velocity range is $\rm V_{\rm
LSR}=1240-1890\ km\ s ^{-1}$, and the resolution is $\rm \Delta V_{\rm
chan}=10.5$ km s$ ^{-1}$.
}

{\bf Fig 2}
{The grid map
of all the $ ^{12}$CO J=3--2 spectra. The temperature scale is $\rm
T_{\rm mb} = -0.2...1.0$ ~K, the velocity range is $\rm V_{\rm
LSR}=1161-1963\ km\ s ^{-1}$, and the resolution is $\rm \Delta V_{\rm
chan}=10.8$ km s$ ^{-1}$.
}

{\bf Fig 3}
{Comparison of the  CO J=3--2 profiles  
 (thick) with the J=1--0 profiles (thin). The former has been re-gridded
 and convolved to an angular resolution of $15''$ for comparison with the
CO J=1--0 data.  The temperature scale is $\rm
T_{\rm mb} = -0.2...1.0$ ~K, and the velocity range is $\rm V_{\rm
LSR}=1240-1890\ km\ s ^{-1}$.
}

{\bf Fig 4}
{(a) The $^{12}$CO J=1--0 contour maps overlaid on the J=3--2 
greyscale map (top). The contour levels are 16 
K \kmps and increment by 16 K \kmpsnospace.
(b) The  derived $r_{31}$ contours and greyscale map (bottom). 
The circle at the lower right corner indicates the resolution (15$''$). 
The contour levels are 0.10, 0.25, 0.4, 0.55, 0.70, 0.85.
}

{\bf Fig 5} 
{The $ ^{13}$CO J=3--2 profiles overlaid on the re-observed $ ^{12}$CO
J=3--2 profiles at the center of the overlap region (the map center in
Figures 1 and 2) and the NGC~4038 nucleus.
}

{\bf Fig 6}
{The $ ^{13}$CO J=2--1  profiles overlaid on
 the $ ^{12}$CO J=2--1 profiles  in the overlap region and the two nuclei.
}

{\bf Fig 7} { Channel maps of (a) $ ^{12}$CO J=1--0 emission (top),
and (b) $ ^{12}$CO J=3--2 emission (bottom).  For both maps, each
velocity channel is 20 \kmpsnospace. The offsets are relative to R.A.=
$12^h01^m54.8^s$, DEC= $-18^o52'55''$ (J2000). The contour levels are
0.1, 0.2, 0.3...0.9 K in $\rm T_{\rm mb}$ units. A K-band NIR image is shown
as the background of the last channel map. The star, triangle and cross 
symbols indicate respectively
the C1, C2 and C3 components.}

{\bf Fig 8}
{The $^{12}$CO J=3--2/1--0 contour maps overlaid on (a)the K-band NIR
image (top) and (b) the ISO 15 \um image (bottom). 
The circles at the lower right corner
indicate the resolution which is 15$''$. The contour levels are 0.10,
0.25, 0.4, 0.55, 0.70, 0.85.  
}

{\bf Fig 9}
{The SCUBA 850 \um contours overlaid on the JCMT CO J=3--2 greyscale map 
(a) before (top) and (b) after (bottom) correcting
for the CO J=3--2 contamination. The circle at the lower right corner 
indicates a  15$''$ beamsize. The contour levels are 35, 58, 81, 104 mJy/beam in (a) and
 22, 35.5, 49, 62.5 mJy/beam in (b).
}

{\bf Fig 10}
{The SCUBA 850 \um  (white) and 450 \um contours (black) overlaid on
the Nobeyama CO J=1--0 map. 
The circles at the lower right corner 
indicate a  15$''$ beamsize. The contour levels are 22, 35.5, 49, 62.5 
mJy/beam for the 850 \um contours and 30, 50, 70, 90 percent of the peak 
for the 450 \um contours.
}


{\bf Fig 11}
{The corrected SCUBA 850 \um  map (contours) overlaid on the HST full color image of the 
``Antennae'' galaxies. The contour levels are 22, 35.5, 49, 62.5 mJy/beam.

{\bf Fig 12}
{The corrected SCUBA 850 \um  map (contours) overlaid on the HST H$_\alpha$ image of the 
``Antennae'' galaxies. The contour levels are 22, 35.5, 49, 62.5 mJy/beam.



\end{document}


{\small
\begin{center} 
{\bf \large Table 1: Integrated intensity ratios of $^{12}$CO J=3--2/1--0 }
\vskip 2.ex
\tabcolsep 16pt
\begin{tabular}{ccccccc}
\hline
\hline
   $\Delta \alpha     \, \, ^a $  &
   $\Delta \delta      \, \, ^a $    &
$I_{10} \, \, ^b $  &
$I_{32} \, \, ^c$  &
$ r_{31} \, \, ^d$ &
$\sigma_{31} \, \, ^e$ 
\\
  $''$      &
  $''$    &
K \kmps   &
K \kmps &
           &
           \\
\hline
-37.0 & 42.5 & 38.2 & 17.4 & 0.46 & 0.12 \\
-37.0 & 51.0 & 29.9 & 33.0 & 1.10 & 0.17 \\
-28.5 & 42.5 & 47.0 & 29.8 & 0.63 & 0.10 \\
-28.5 & 51.0 & 66.8 & 62.3 & 0.93 & 0.08 \\
-28.5 & 59.5 & 73.9 & 46.0 & 0.62 & 0.06 \\
-20.0 & -17.0 & 77.2 & 46.5 & 0.60 & 0.05 \\
-20.0 & -25.5 & 41.2 & 32.9 & 0.78 & 0.11 \\
-20.0 & -8.5 & 40.8 & 26.0 & 0.62 & 0.09 \\
-20.0 & 34.0 & 39.8 & 12.8 & 0.32 & 0.13 \\
-20.0 & 42.5 & 47.0 & 28.4 & 0.60 & 0.10 \\
-20.0 & 51.0 & 61.4 & 48.8 & 0.80 & 0.09 \\
-20.0 & 59.5 & 47.9 & 32.9 & 0.69 & 0.10 \\
-11.5 & -17.0 & 36.1 & 42.6 & 1.18 & 0.14 \\
-11.5 & -8.5 & 48.6 & 40.3 & 0.85 & 0.08 \\
-11.5 & 0.0 & 72.1 & 32.5 & 0.45 & 0.04 \\
-11.5 & 8.5 & 56.6 & 24.6 & 0.43 & 0.05 \\
-11.5 & 17.0 & 26.8 & 17.1 & 0.64 & 0.12 \\
-11.5 & 25.5 & 25.9 & 12.3 & 0.48 & 0.17 \\
-11.5 & 34.0 & 30.7 & 13.2 & 0.43 & 0.16 \\
-11.5 & 42.5 & 23.8 & 15.4 & 0.65 & 0.23 \\
-11.5 & 51.0 & 44.2 & 15.2 & 0.34 & 0.11 \\
-3.0 & -17.0 & 51.6 & 43.8 & 0.85 & 0.10 \\
-3.0 & -8.5 & 93.8 & 69.2 & 0.74 & 0.04 \\
-3.0 & 0.0 & 98.5 & 74.4 & 0.76 & 0.04 \\
-3.0 & 8.5 & 80.5 & 56.8 & 0.70 & 0.04 \\
-3.0 & 17.0 & 70.6 & 30.4 & 0.43 & 0.05 \\
-3.0 & 25.5 & 52.1 & 22.9 & 0.44 & 0.07 \\
-3.0 & 34.0 & 25.1 & 14.7 & 0.59 & 0.21 \\
5.5 &  17.0 & 72.3 & 32.8 & 0.45 & 0.04 \\
5.5 &  25.5 & 34.8 & 18.0 & 0.52 & 0.12 \\
5.5 & -17.0 & 18.8 & 29.6 & 1.57 & 0.29 \\
5.5 & -8.5 & 74.6 & 62.0 & 0.83 & 0.05 \\
5.5 & 0.0 & 110.0 & 81.0 & 0.74 & 0.03 \\
5.5 & 8.5 & 107.5 & 67.4 & 0.63 & 0.03 \\
14.0 & -17.0 & 36.5 & 10.9 & 0.30 & 0.10 \\
14.0 & -8.5 & 51.4 & 23.3 & 0.45 & 0.07 \\
14.0 & 0.0 & 61.7 & 42.0 & 0.68 & 0.06 \\
14.0 & 8.5 & 60.4 & 42.2 & 0.70 & 0.06 \\
14.0 & 17.0 & 60.5 & 23.3 & 0.39 & 0.05 \\
\hline
\end{tabular}
\end{center}
\begin{small}
\noindent
\noindent
$ ^{\rm a}$   $\Delta \alpha$ and   $\Delta \delta$  are the R.A and DEC offsets from the map center R.A.= $12^h01^m54.8^s$, \\ DEC= $-18^o52'55''$ (J2000). \\
$ ^{\rm b}$ The integrated intensity ($ T_{\rm mb}$) of CO J=1--0.\\
$ ^{\rm c}$ The integrated intensity ($ T_{\rm mb}$) of CO J=3--2, which
have been convolved from 14$''$ to 15$''$ and re-gridded to match the CO J=1--0 data. \\
$ ^{\rm d}$ The average integrated intensity ratios in a 15$''$ beam. \\
$ ^{\rm e}$ The uncertainty in the $r_{31}$ ratio.
\end{small}
}

\clearpage

{\small
\begin{center} 
{\large \bf Table 2: Observed line ratios $^a$}
\vskip 2.ex
\tabcolsep 16pt
\begin{tabular}{lccccc}
\hline
\hline
Position  &
$R_{21}$ $^b$ &
$r_{21}$  &
$R_{32}$   &
$r_{31}$  &
$^{13}r_{32}$ \\
\hline
N4038   &  13 $\pm$ 3  &1.01$\pm$ 0.1  &   19 $\pm$  4  & 0.91 $\pm 0.1$     & 0.6$\pm$ 0.2\\
N4039   &  16 $\pm$ 4  &1.19$\pm$ 0.1 &   --            & 0.73 $\pm 0.08$   & -- \\
Overlap &              &             &         &                &     \\  
(0,0)   &  26 $\pm$ 5  &1.04$\pm$ 0.1  &  25 $\pm$ 8    &  0.78 $\pm 0.09$    & 0.77$\pm$0.23\\
(0,10)  &  29 $\pm$ 5  &1.11$\pm$ 0.1  &        --             & 0.67 $\pm 0.08$    & --\\
(0,-10) &  18 $\pm$ 4  &1.06$\pm$ 0.1  &        --             & 0.87 $\pm 0.1$    & --\\
\hline
\end{tabular}
\end{center}
\begin{small}
\noindent
\noindent
$ ^{\rm  a}$ The average line intensity ratios in  a 20$''$ beam. The intensities with a higher resolution have been convolved
to 20$''$. \\
$ ^{\rm  b}$ The quantities $R_{21}$, $r_{21}$, $R_{32}$, $r_{31}$ and $^{13}r_{32}$ are defined in Section 6.

\vspace*{0.3cm}
\end{small}
}

\clearpage

{\small
\begin{center} 
{\large \bf Table 3: One-component model fitting result}
\vskip 2.ex
\tabcolsep 3pt
\begin{tabular}{lcccccccc}
\hline
\hline
{Position} &
{\nh2} &
{\Tk} &
{X$_{CO}$/(dV/dR)} &
$N_{CO}/\Delta V$ &
$\tau_{10}$  &
$f$             &
{ \1213CO } &
{$\chi   $} 
\\
{} &
{$10^3$ \cm3} &
{K} &
{$ 10^{-6}$ pc (km/s)$^{-1}$} &
{$10^{16}$ cm$^{-2}$ (km/s)$^{-1}$} &
{ } &
{ } &
{} &
{}
\\ 
\hline  
N4038    &  3.5      &  43     &    3.2 & 3.4    &  0.70   &0.050&   40    &   1.2\\ 
         &  1-- 4   &  35--160 &  2-- 35&2.5--10& 0.4--1.1&--   &   40--45 &   $<$2 \\ 
N4039    &  3.2      &  35     &    2.3 & 2.5    &  0.62   &0.018&  40     &   0.8 \\
         &   1-- 10  & 20--130 &  1--10 &2.2--3.1& 0.3--1.0 &--   &  40     &   $<$2 \\ 
Overlap  &           &         &        &        &         &     &         &       \\
(0,0)    &   3.0     & 36      &   3.0  & 2.5    &  0.90   &0.051&   70    &   2.2 \\ 
         &   1-- 6   & 28--160 & 1--10  &1.8--3.1& 0.3--1.0 &--   & 40--200  &   $<$3  \\ 
(0,10)   &   3.2     &  33     &   1.0  & 0.98   & 0.33    &0.061&  60     &   1.2 \\
         &   1-- 5   & 25--130 &  1--10 &1.5--3.1& 0.2--1.0 &--   &  $>$40  &   $<$2 \\ 
(0,-10)  &   10      &  25     &   1    &  3.1   &  1.3    &0.030&  70     &   0.9 \\ 
         &   1-- 20  & 20--130 &  1--10 &3.1--6.2& 1.0--1.5 & --  &  $>$40  &   $<$2 \\ 
\hline 
\end{tabular}
\end{center}
}
\noindent
Note: for each position, the first row is the best fit parameters and
the second row is the estimated range within the observed uncertainty.
\vspace*{2cm}
\clearpage

\centerline{ \bf Table 4:\hspace*{0.5cm}Two-component modeling for the overlap region}
\begin{center}
\begin{tabular}{| c | c | c | c | c |} \hline\hline
$[\frac{^{12}CO}{^{13}CO}]$ &
LVG parameters $ ^{\rm a}$ &
Line ratios    $ ^{\rm b}$   & 
$\rm N_{CO}/\Delta V \, ^{\rm c}$   & 
 $\chi _{\rm min} \,  ^{\rm d}$  
\\ 
                                                     &
$\rm  T_{\rm   kin}$, $\rm n(H_2)$,  $\Lambda$, $f $ &  
$\rm r_{21}$, $\rm  r_{31}$,  $\rm R_{21},  R_{32}$  &  
$  cm^{-2} /$ \kmps &
                           \\ 
\hline
70 &(A) 36, $\rm 3.2 x 10^3$, $\rm 3 x 10^{-6}$, 0.051   & 1.06, 0.80, 23, 28  & 1.9 $\times 10^{15}$ & 0.9\\  
   &(B) 90, $\rm 1.0 x 10^5$, $\rm 3 x 10^{-6}$, 0.00033 & (1.04, 0.78, 26, 25)&     & \\ \hline
70 &(A) 120, $\rm 1.0 x 10^3$, $\rm 1 x 10^{-5}$, 0.044  & 1.07,  0.78, 24, 25  & 1.8 $\times 10^{15}$  & 0.6  \\
   &(B) 90, $\rm 1.0 x 10^5$, $\rm 3 x 10^{-6}$, 0.0005  & (1.04, 0.78, 26, 25)&     &\\ \hline
70 &(A) 120, $\rm 1.0 x 10^3$, $\rm 1 x 10^{-5}$, 0.044  &1.07, 0.79, 23, 29   & 1.6 $\times 10^{15}$  & 1.1 \\
   &(B) 42, $\rm 3.2 x 10^4$, $\rm 1 x 10^{-6}$, 0.0022  & (1.04, 0.78, 26, 25)&     & \\ \hline
40 &(A) 120, $\rm 1.0 x 10^3$, $\rm 3 x 10^{-6}$, 0.10   & 1.13, 0.70, 23, 27  & 1.5  $\times 10^{15}$  & 1.6 \\   
   &(B) 220, $\rm 1.0 x 10^5$, $\rm 3 x 10^{-6}$, 0.00060 & (1.04, 0.78, 26, 25)&     & \\ \hline
\end{tabular}
\end{center}
\begin{small}
\noindent
$ ^{\rm  a}$ LVG parameters for component A (first row) and B (second row). $f$ is the filling 
factor of each component.
\hspace*{0.2cm} The units are the same ones  used in Table 3. 
\\  
\noindent
$ ^{\rm  b}$ The line ratios estimated  for the combination of two components.
 The  ones in the parentheses  are the  observed values. 
\\
\noindent
$ ^{\rm  c}$ $\rm  N_{CO}$ is the total column density  combining the components using 
the formula: \\
$ N_{CO}/\Delta V = [n(H_2) \, X_{CO}/(dV/dR)\,  f ]_A + [n(H_2) \, X_{CO}/(dV/dR)\,  f ]_B $ 
\\
\noindent
$ ^{\rm d}$ The $\chi  _{\rm  min}$ parameter
 characterizes the goodness of the fit (see Table 3)
\end{small}
\clearpage

{\small
\begin{center} 
{ \bf Table 5: Comparison of SCUBA 850 \um flux and derived dust mass 
              before and after correcting the CO J=3--2 contamination}
\vskip 2.ex
\tabcolsep 6pt
\begin{tabular}{lccccc}
\hline
\hline
{Position} &
{$S_{450}$ $^a$} &
{$S_{850}$ (Haas) $^b$} &
{$S_{850}$ (revised) $^c$ } &
{$M_{dust}$ (Haas) $^d$} &
{$M_{dust} $(revised) $^e$} 
\\
{ } &
{Jy} &
{mJy} &
{mJy} &
{$ 10^7 \ms$} &
{$ 10^7 \ms$} 
\\
\hline
NGC4038    &  1.69  &    250  & 166.8  & 2.8      & 0.5\\
NGC4039     &  1.06  & 150     & $ <$ 78&1.7       & $<$0.3\\
K1       &  1.33  & 190     &  83.8  &$>$2.2    &0.25 \\
K2       & $<$ 0.90& 170    &  106   &$>$3.8    &0.30 \\
\hline 
\end{tabular}
\end{center}
\noindent
\noindent
$ ^{\rm  a}$  The SCUBA 450 \um flux from Table 1 of  Haas \etal  (2000). 
\\
\noindent
$ ^{\rm  b}$  The SCUBA 850 \um flux from Table 1 of  Haas \etal  (2000).
\\ 
\noindent
$ ^{\rm  c}$  The revised 850 \um continuum flux after correcting for CO J=3--2 
contamination.  See text for details. 
\\
\noindent
$ ^{\rm  d}$  The dust mass estimated by  Haas \etal  (2000).
\\
\noindent
$ ^{\rm e}$ The dust mass estimated from the real 850 \um continuum
flux. The dust model was the same as that of Haas \etal  (2000). 
See text for details.
\vspace*{0.3cm}
}
\clearpage